\documentclass[12pt]{article}
\usepackage{aaspp4}
\usepackage{graphicx}
\usepackage{epsfig}
\usepackage{subfigure}
\usepackage{hyperref}
\usepackage{multicol}
\let\oldfootsep=\footnotesep
\def\simlt{\hbox{ \rlap{\raise 0.425ex\hbox{$<$}}\lower 0.65ex\hbox{$\sim$} }}
\def\simgt{\hbox{ \rlap{\raise 0.425ex\hbox{$>$}}\lower 0.65ex\hbox{$\sim$} }}

\def\msun {M_\odot}
\def\mearth {M_\oplus}

\def\etal{{et al.}}

\def\msun{M_\odot}

\def\leaderfill{\leaders\hbox to 1em{\hss-\hss}\hfill}

\def\lsim{\mathrel{\mathpalette\@versim<}}
\def\gsim{\mathrel{\mathpalette\@versim>}}
\def\@versim#1#2{\lower0.2ex\vbox{\baselineskip\z@skip\lineskip\z@skip
  \lineskiplimit\z@\ialign{$\m@th#1\hfil##\hfil$\crcr#2\crcr\sim\crcr}}}

\def\spose#1{\hbox to 0pt{#1\hss}}
\def\simlt{\mathrel{\spose{\lower 3pt\hbox{$\mathchar"218$}}
     \raise 2.0pt\hbox{$\mathchar"13C$}}}
\def\simgt{\mathrel{\spose{\lower 3pt\hbox{$\mathchar"218$}}
     \raise 2.0pt\hbox{$\mathchar"13E$}}}

\begin{document}

\rightskip = 0.0in plus 1em
\pagestyle{plain}

\centerline{\large \bf  The WFIRST Exoplanet Microlensing Survey }
\vskip 0.1in

\centerline{David P.~Bennett$^1$ (301-286-5473, david.bennett@nasa.gov),}
\noindent  
Rachel Akeson$^2$,
Jay Anderson$^3$,
Lee Armus$^2$,
Etienne Bachelet$^4$,
Vanessa Bailey$^5$,
Thomas Barclay$^1$,
Richard Barry$^1$,
Jean-Phillipe Beaulieu$^6$,
Andrea Bellini$^3$,
Dominic J.~Benford$^7$,
Aparna Bhattacharya$^1$, 
Padi Boyd$^1$,
Valerio Bozza$^8$,
Sebastiano Calchi Novati$^2$,
Kenneth Carpenter$^1$,
Arnaud Cassan$^9$,
David Ciardi$^2$,
Andrew Cole$^6$,
Knicole Colon$^1$,
Christian Coutures$^9$,
Martin Dominik$^{10}$,
Pascal Fouqu\'e$^{11}$
Kevin Grady$^1$,
Tyler Groff$^1$,
Calen B.~Henderson$^2$,
Keith Horne$^{10}$,
Christopher Gelino$^2$,
Dawn Gelino$^2$,
Jason Kalirai$^3$,
Stephen Kane$^{12}$,
N.~Jeremy Kasdin$^{13}$,
Jeffrey Kruk$^1$,
Seppo Laine$^2$,
Michiel Lambrechts$^{14}$,
Luigi Mancini$^{15}$,
Avi Mandell$^1$,
Sangeeta Malhotra$^1$,
Shude Mao$^{16}$,
Michael McElwain$^1$,
Bertrand Mennesson$^5$,
Tiffany Meshkat$^2$,
Leonidas Moustakas$^5$,
Jose A.~Mu\~noz$^{17}$,
David Nataf$^{18}$,
Roberta Paladini$^2$,
Ilaria Pascucci$^{19}$,
Matthew Penny$^{20}$,
Radek Poleski$^{20}$,
Elisa Quintana$^1$,
Cl\'ement Ranc$^1$,
Nicholas Rattenbury$^{21}$,
James Rhodes$^1$,
Jason D.~Rhodes$^5$,
Maxime Rizzo$^1$, 
Aki Roberge$^1$,
Leslie Rogers$^{22}$,
Kailash C.~Sahu$^3$,
Joshua Schlieder$^1$,
Sara Seager$^{23}$,
Yossi Shvartzvald$^5$,
R\'emi Soummer$^3$, 
David Spergel$^{13}$,
Keivan G.~Stassun$^{24}$,
Rachel Street$^4$, 
Takahiro Sumi$^{25}$, 
Daisuke Suzuki$^{26}$,
John Trauger$^5$,
Roeland van der Marel$^3$,
Benjamin F. Williams$^{27}$,
Edward J.~Wollack$^1$,
Jennifer Yee$^{28}$,
Atsunori Yonehara$^{29}$, and
Neil Zimmerman$^1$

\vspace{0.3cm}
\begin{multicols}{2}
\noindent $^1$NASA Goddard Space Flight Center \\
$^2$IPAC/Caltech \\
$^3$Space Telescope Science Institute \\
$^4$Las Cumbres Observatory \\
$^5$Jet Propulsion Laboratory \\
$^6$University of Tasmania, Australia \\
$^7$NASA Headquarters \\
$^8$University of Salerno, Italy \\
$^9$Institut d'Astrophysique de Paris, France \\
$^{10}$University of St.Andrews, Scotland, UK \\
$^{11}$Canada France Hawaii Telescope Corp. \\
$^{12}$University of California, Riverside \\
$^{13}$Princeton University \\
$^{14}$Lund University, Sweden \\
$^{15}$University of Rome Tor Vergata, Italy \\
$^{16}$Tsinghua University, China \\
$^{17}$University of Valencia, Spain \\
$^{18}$Johns Hopkins University \\
$^{19}$University of Arizona \\
$^{20}$Ohio State University \\
$^{21}$University of Auckland, New Zealand \\
$^{22}$University of Chicago \\
$^{23}$Massachusetts Institute of Technology \\
$^{24}$Vanderbilt University \\
$^{25}$Osaka University, Japan \\
$^{26}$ISAS, JAXA, Japan \\
$^{27}$University of Washington \\
$^{28}$Harvard-Smithsonian CfA \\
$^{29}$Kyoto Sangyo University, Japan \\
\end{multicols}

\centerline{Submitted to the Committee on an Exoplanet Science Strategy}

\pagebreak

\section{Scientific Rationale}
\label{sec-rationale}
\vspace{-0.1cm}

The Wide Field Infrared Survey Telescope (WFIRST) was the top ranked large space mission
in the 2010 New Worlds, New Horizons decadal survey, and it was formed by merging
the science programs of 3 different mission concepts, including the Microlensing Planet
Finder (MPF) concept (Bennett \etal\ 2010). The WFIRST science program (Spergel \etal\ 2015)
consists of a general observer program, a wavefront controlled technology program, and two targeted 
science programs: a program to study
dark energy, and a statistical census of exoplanets with a microlensing survey, which uses 
nearly one quarter of WFIRST's observing time in the current design reference mission. The
New Worlds, New Horizons (decadal survey) midterm assessment summarizes the science 
case for the WFIRST exoplanet microlensing survey with this statement:
``WFIRST's microlensing census of planets beyond 1 AU will perfectly complement Kepler's 
census of compact systems, and WFIRST will also be able to detect free-floating planets 
unbound from their parent stars\rlap."

\begin{figure}[!ht]
\centerline{\includegraphics[width=5.0in]{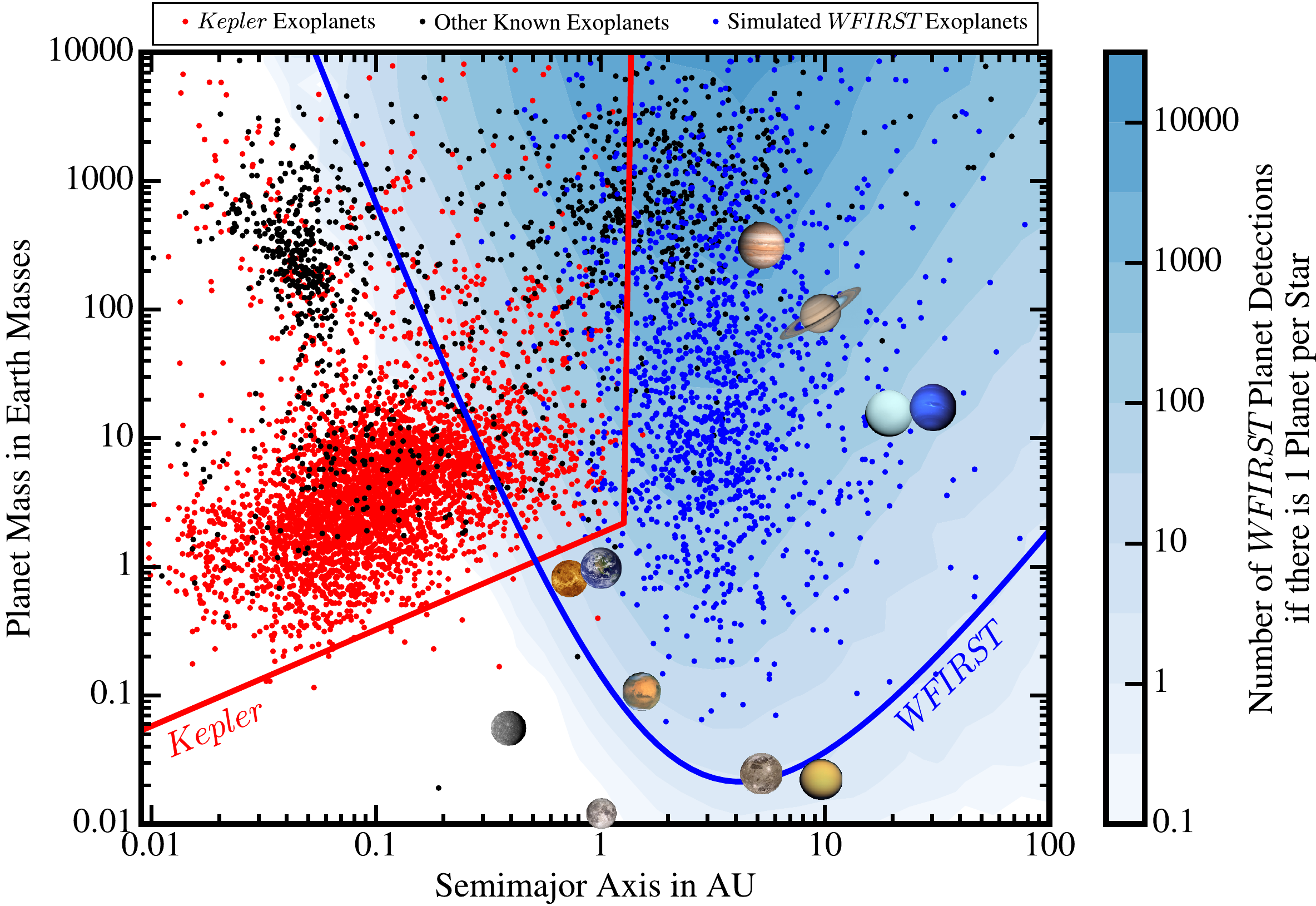}}
\vspace{-0.4cm}
\caption{Comparison of the WFIRST design sensitivity to that of Kepler in the planet 
mass-semimajor axis plane (Penny \etal, in preparation).
The red line shows the approximate 50\% completeness limit of Kepler
(Burke \etal\ 2015). Blue shading shows the number of WFIRST planet detections during the 
mission if there is one planet per star at a given mass and semimajor axis. The thick blue line is 
the contour for 3 detections. Red dots show Kepler candidate and confirmed planets, black dots show 
all other known planets. Blue dots show a simulated realization of the planets detected by the 
WFIRST microlensing survey, assuming a fiducial planet mass function. Solar system bodies are shown 
by their images, including the moons Ganymede, Titan, and The Moon at the semi-major axis of their hosts.
\label{fig-mass_sma}}
\end{figure}

\begin{figure}[h]
 \begin{minipage}[t]{0.50\textwidth}
  \includegraphics[width=\textwidth]{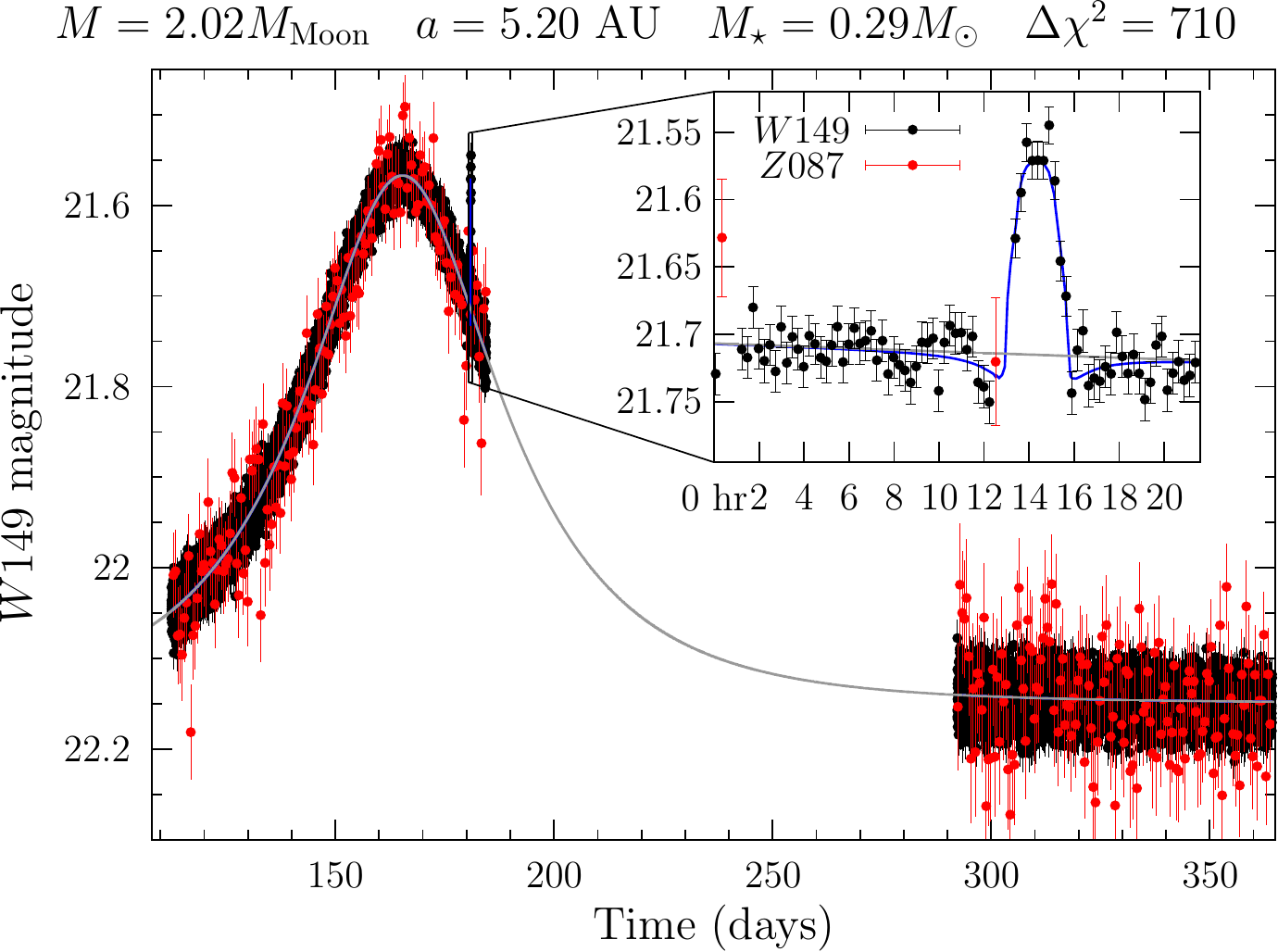}
 \end{minipage}\hfill
 \begin{minipage}[b]{0.45\textwidth}
\caption{Example WFIRST lightcurve of a $0.025\mearth$ (2 lunar-mass) bound planet 
detection. Black and red data points are in the wide W149 and Z087 filters,
respectively. The blue curve shows the underlying ÒtrueÓ lightcurve and the grey line shows
 the best fit single lens light curve.}\label{fig-WFIRST_lc}
 \end{minipage}
\end{figure}

WFIRST is needed to detect low-mass planets at orbital separations $\simgt 1\,$AU,
complementing Kepler's sensitivity to short period planets.
Fig.~\ref{fig-mass_sma} shows that WFIRST is sensitive to planets in these orbits
with masses two orders of magnitude smaller than what is achievable with other methods.
It indicates that WFIRST will
be sensitive to analogs to all the planets in our Solar System, except for Mercury, and its 
sensitivity reaches down to very low mass planets as indicated by the light curve from 
a WFIRST simulation shown in Fig.~\ref{fig-WFIRST_lc} (Penny \etal\ in preparation). 
Even very low mass planets can be detected with strong signals. WFIRST is also sensitive to 
free-floating planets down to the mass of Mars that have been ejected from the planetary systems of their
birth, which should provide important clues the processes of planet formation.

The ultimate goal of NASA's exoplanet program is to search for signs of life on other 
planets, but this requires an understanding of the planet formation process. Crucial 
details, such as the delivery of water to Earth, are likely to depend on the details of
the planet formation process (Raymond \etal\ 2007). So, the presence of an
Earth-size planet in the habitable zone may not be a sufficient condition to allow life to develop.
On the other hand, our understanding
of planetary habitability is currently rather primitive, so it could be that planets
very different from Earth could have conditions that are suitable for life (Seager 2013).
Thus, gaining a better understanding of the basic properties of exoplanets and their 
formation mechanisms is a critical part of NASA's search for life outside the Solar System.
It has proved to be quite challenging to develop a detailed understanding of planet formation
process, as it seems to involve a wide range of physical processes operating over a huge
range of length scales. It is fair to say that planet formation theory has yet to provide
any detailed predictions that have been confirmed, and that progress in the theoretical
understanding of planet formation must closely follow observational discoveries.
Thus, WFIRST's exoplanet microlensing survey is a crucial element of NASA's 
exoplanet program.


\begin{figure}[h]
 \begin{minipage}[t]{0.59\textwidth}
  \includegraphics[width=\textwidth]{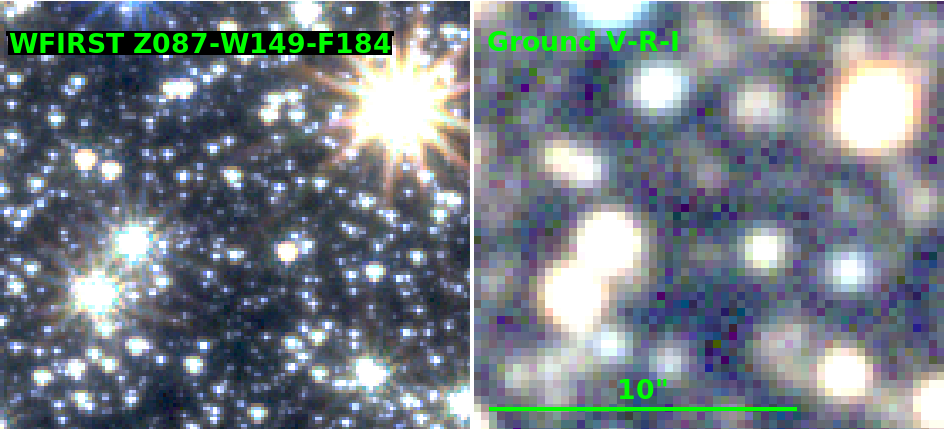}
 \end{minipage}\hfill
 \begin{minipage}[b]{0.40\textwidth}
\caption{Simulated color images of a WFIRST microlensing survey field,
as observed by WFIRST in the IR and a optical ground-based telescope. 
Only the blue foreground disk stars and bulge giants are resolvable in the ground-based
images on the right, while the much more numerous main sequence stars are
resolved in the WFIRST images.}\label{fig-WFIRST_img}
 \end{minipage}
\end{figure}

Ground-based microlensing is already playing a crucial role in our understanding
of planet populations. Suzuki \etal\ (2016) have found a break
and likely peak in the exoplanet mass ratio function at a mass ratio of $q\sim 10^{-4}$,
and other ongoing surveys promise substantially improved statistics. 
Fully characterizing the mass-ratio and mass functions can only be done from
space. Ground-based microlensing rapidly runs out of sensitivity to
planets below the observed break in the mass ratio function. 

Signals from the smallest planets (mass $\simlt 1\mearth$)
are largely washed out by finite-source effects (Bennett \& Rhie 1996),
especially if they orbit inside the Einstein radius, which is
typically 2-3$\,$AU. These effects are minimized for dwarf source stars,
but the bulge main sequence stars that comprise the vast majority of
microlensing source stars are not resolved in the ground-based
images. This means that the ground-based photometry of these stars is
imprecise due to blending with other unresolved stars and that the
microlensing signal is diluted. WFIRST's vastly improved resolution
(Fig.~\ref{fig-WFIRST_img}) will resolve these issues.

This improved resolution will also enable the characterization of the host stars by allowing 
direct measurements of their flux and the vector lens-source relative proper motion, 
$\mbox{\boldmath$\mu$}_{\rm rel}$ (Bennett \etal\ 2006, 2015)
WFIRST will also combine relative proper motion measurements with partial microlensing
parallax measurements to measure masses more directly
(Gould 2014). Combining the mass-distance relationships from 
microlens parallax and flux
measurements also determines the mass of the lens (Yee 2015). Measurements
of $\mbox{\boldmath$\mu$}_{\rm rel}$ also ensure the the measured light can
be correctly identified with the exoplanet host star (Bhattacharya \etal\ 2017; Koshimoto \etal\ 2017).

\vskip -0.1cm
\begin{figure}[!ht]
\centerline{\includegraphics[width=6.0in]{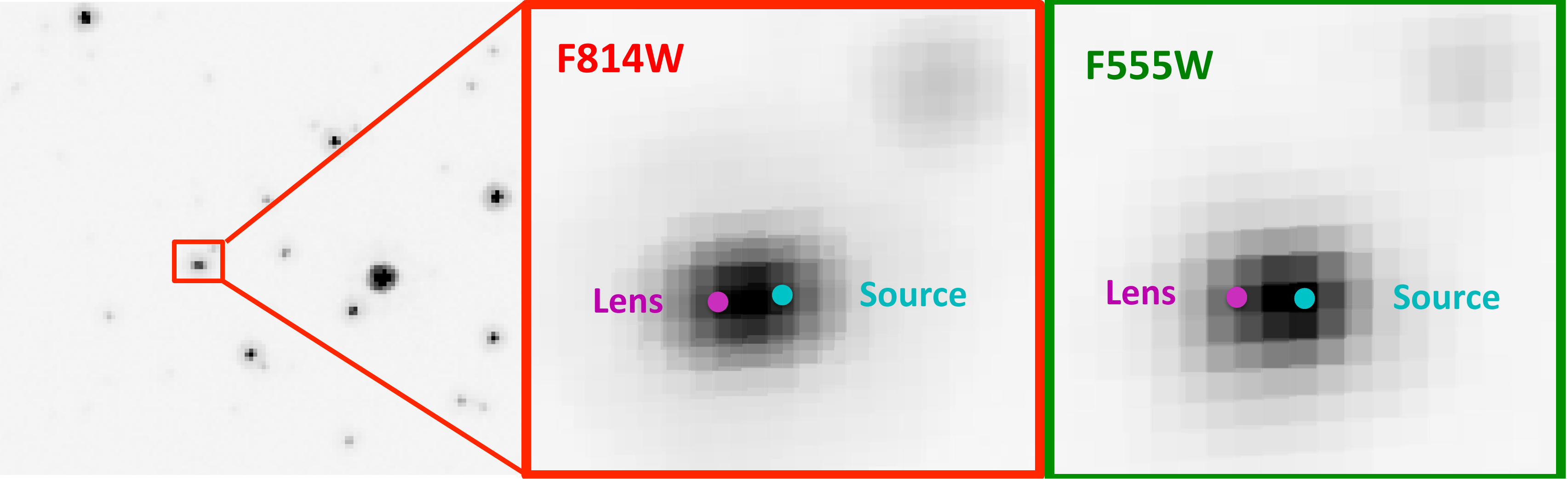}}
\vspace{-0.4cm}
\caption{Stacked, dithered HST/WFC3 images of the OGLE-2005-BLG-169 source and lens star,
separated by 50$\,$mas in images taken 6.5 years after the microlensing event. 
The fit to microlensing light curve constraints and mass-magnitude relations yields the masses of
the host (or lens) star and its planet of $M_L= 0.69 \pm 0.02\msun$
and $m_p = 14.1\pm 0.9\mearth$ (Bennett \etal\ 2015).
\label{fig-mass_hst}}
\end{figure}

Fig.~\ref{fig-mass_hst} shows HST images that reveal the lens (and planet host) star
separating from the source star at the predicted $\mu_{\rm rel}$ value
for planetary microlensing event OGLE-2005-BLG-169. High angular
resolution follow-up observations from space or narrow field adaptive
optics systems are needed to measure such effects for events found
from the ground, but WFIRST will measure such effects directly. With
$\sim 40,000$ images per star, WFIRST will have the S/N to measure
lens-source separations much smaller than seen in Fig.~\ref{fig-mass_hst}.

\section{Developments Since the Decadal Survey}
\label{sec-since}
\vspace{-0.1cm}


The WFIRST mission has undergone some changes since the Decadal Survey. The main
change has been the move from a 1.3-1.5m telescope to a 2.4m aperture telescope that
was gifted to NASA from another government agency. This
has opened up some opportunities for WFIRST. It has enabled a technology demonstration
wavefront controlled coronagraph to be added to WFIRST, and it allows the dark energy 
studies to go much deeper in contrast to ESA's Euclid mission. The improvements for the
exoplanet microlensing survey have been more modest, because the smaller field-of-view
and larger slew times than previous designs counteract some of the gains in photometric 
precision. However, the significant improvement in angular resolution adds a great deal of
margin to the lens-source relative proper motion ($\mu_{\rm rel}$) measurements that are part of the 
primary WFIRST mass measurement method, illustrated in Fig.~\ref{fig-mass_hst}.

\section{Simultaneous Ground-based Surveys}
\label{sec-simult}
\vspace{-0.1cm}

\begin{figure}[h]
 \begin{minipage}[t]{0.49\textwidth}
  \includegraphics[width=\textwidth]{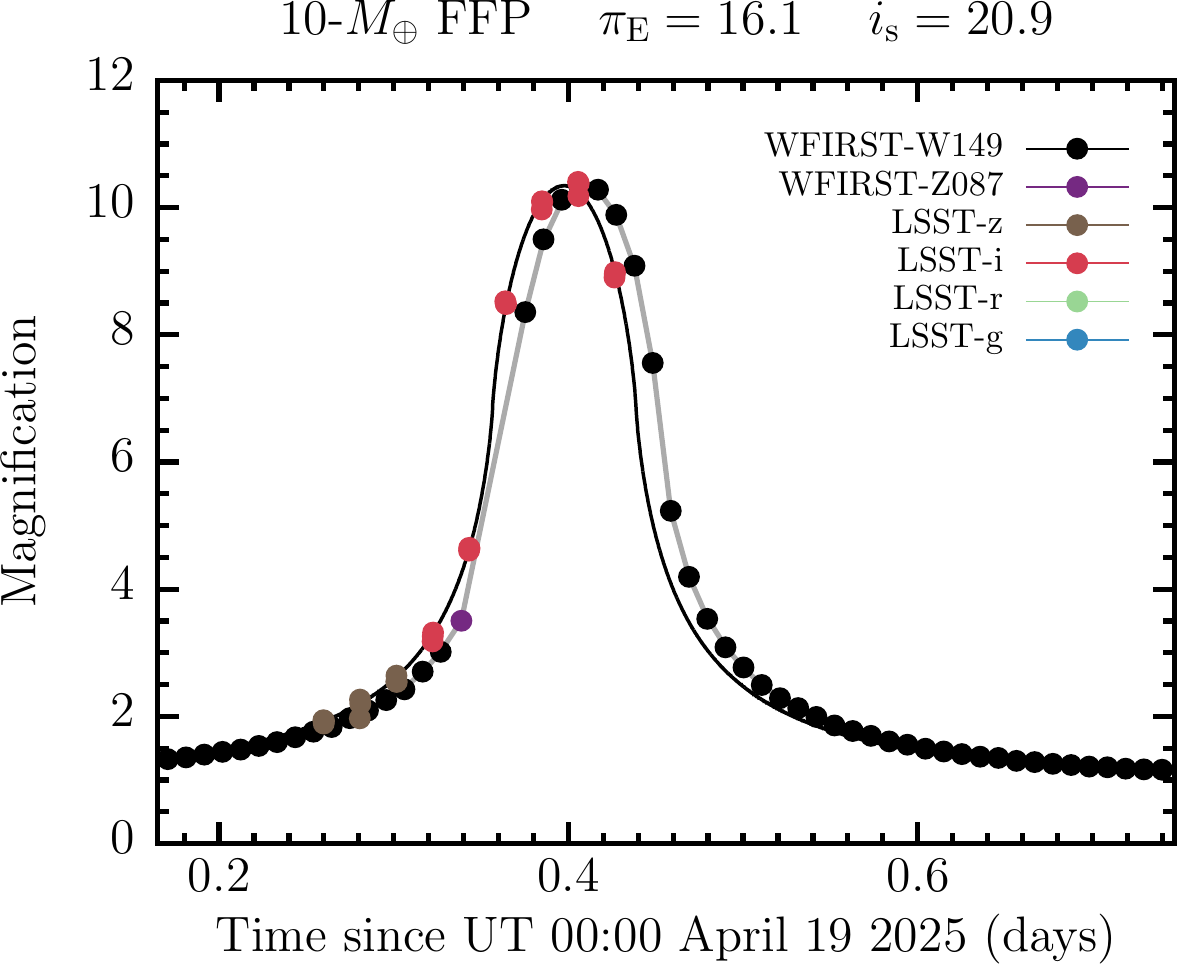}
 \end{minipage}\hfill
 \begin{minipage}[b]{0.46\textwidth}
\caption{Simulated light curve of a free-floating planet of $10\mearth$ observed
simultaneously by WFIRST and LSST. The shift in the time of peak magnification
between the two light curves is due to the microlensing parallax effect. When combined
with the source radius crossing time, $t_*$, that is also measured from this 
light curve, the microlensing parallax measurement will yield the host star
mass.}\label{fig-ffp_WFIRST_LSST}
 \end{minipage}
\end{figure}

The primary WFIRST exoplanet host star and planet mass measurement method involves 
measuring the brightness and relative proper motion of the host star, so this method does not apply to
planets without a host star.
The same is true for planets hosted by brown dwarfs or white dwarfs. Fortunately, for 
events caused by low-mass free-floating or bound planets, there is a method that
can determine the lens masses for events that can be observed from the ground.
This method uses finite source effects (which yield the angular Einstein radius, $\theta_E$)
combined with a measurement of the microlensing parallax effect
(as seen in Fig.~\ref{fig-ffp_WFIRST_LSST}) between Earth and WFIRST's L2 orbit.
This combination yields the lens mass (Yee 2013), and this is the only method to 
determine the masses of free-floating planets. This method is only expected to yield
mass measurements for a fraction of free-floating planets, but this will be important,
since it will be our only method to check if our statistical methods to estimate the
mass distribution of free-floating planets might be subject to systematic errors due
uncertainties in their spatial or velocity distribution.

These parallax observations can only be obtained through
simultaneous observations of free-floating planets from the ground and
WFIRST, because the
WFIRST data are not transmitted to the Earth in time for prompt alerts to observe with
follow-up telescopes. There are several planned or existing ground-based telescopes
that could be used for such a ground-based survey, and they would mostly make 
microlensing parallax measurements of different events. So, it would be beneficial 
have several of these simultaneous ground-based surveys. The Large Synoptic Survey 
Telescope (LSST) is one option that could detect relatively faint events (Marshall \etal\ 2017).
The Japanese-American-South African PRIME infrared telescope that is now being 
developed for deployment in South Africa for WFIRST precursor observations (Bennett \etal\ 2018)
would cover a different longitude and would be less affected by high extinction. Another attractive
option would be the Korean Microlensing Telescope Network (KMTNet) (Kim \etal\ 2016) with 
telescopes on 3 Southern continents. A simultaneous WFIRST-KMTNet survey might serve
as a possible Korean contribution to join the WFIRST Project.

\section{Conclusion}
\label{sec-conclude}
\vspace{-0.1cm}
The WFIRST exoplanet microlensing survey is the unique program that can extend the
Kepler census of short period planets to the cooler, long period planets, like seven of the
eight planets in our own Solar System. It will provide fundamental information on the 
properties of exoplanet systems that will be crucial for understanding planet formation
and ultimately the search for life outside the Solar System.

\nopagebreak
\begin{multicols}{2}
{\noindent \bf References} \\
Bennett, D.P.~\& Rhie, S.\ 1996, ApJ, 472, 660 \\
Bennett, D.P.,  et al.\ 2006,  ApJL, 647, L171 \\
Bennett, D.P., et al.\ 2010, arXiv:1012.4486 \\
Bennett, D.P., et al.\ 2015, ApJ, 808, 169 \\
Bennett, D.P., et al.\ 2018, SAG-11 Report update to  CESS \\
Bhattacharya, A., et al.\ 2017, AJ, 154, 59 \\
Burke, C.J., \etal\ 2015, ApJ, 809, 8 \\
Calchi Novati, S., \etal\ 2018, AJ, submitted (arXiv:1801.10586) \\
Gould, A., 2014, KAS, 47, 215 \\
Kim, S.-L., \etal, 2016, JKAS, 49, 37 \\
Koshimoto, N., et al.\ 2017, AJ, 154, 3\\
Marshall, P., \etal\ 2017, arXiv:1708.04058 \\
Raymond, S.~N., Quinn, T., \& Lunine, J.~I.\ 2007, Astrobiology, 7, 66 \\
Seager, S., 2013, Science, 340, 577 \\
Spergel, D. \etal\ 2015, arXiv.org:1503.03757 \\
Yee, J.C., 2013, ApJL, 770, L31 \\
Yee, J.C., 2015, ApJL, 814, L11 \\
Yee, J.C., \etal\ 2018, CESS white paper 
\end{multicols}

\end{document}